# Ph3pyWF: An automated workflow software package for ceramic lattice thermal conductivity calculation

Kerui Lai, Yuxuan Wang

*McGill University, Montreal, Quebec, H3A 0C5, Canada*

## Abstract

This paper presents a Python software package, Ph3pyWF, providing a more convenient platform to realize high-throughput analysis of lattice thermal conductivity for ceramic materials. The interface of Ph3pyWF is friendly for users with different needs and from different level. For novices of interests, the inputs can be quite simple, just with initial unit cell structure. Other parameters would be automatically filled. And for expert-level researchers with varied requirements, plenty of procedure parameters can be customized. The core concept of Ph3pyWF is to build a data exchange and task management system with high efficiency. The design details of the Ph3pyWF will be clarified in this paper and following with a few typical oxide ceramics examples to demonstrate the applicability of this software package.

## Program summary

Program title: Ph3pyWF

Licensing provisions: MIT

Programming language: Python 3

External routines/libraries: Phonopy, Phono3py, Atomate, FireWorks, pymatgen

Nature of problem:

The calculation of lattice thermal conductivity using single-mode relaxation-time approximation and the linearized phonon Boltzmann equation from first-principles method requires a large number of inter-dependent subprocesses. Manually executing and managing such collection of subprocesses is inefficient and error prone.

Solution method:

Employing scientific workflow framework to automate the lattice thermal conductivity calculation process. Providing a near-turn-key solution with simpler management interface to users.

# Introduction

Ceramic with low thermal conductivity can be used in many fields, such as thermal barrier coating in aerospace and power plant [1] and field of thermoelectric material (function can be clarified, like heat block and increase thermoelectric figure of merit). And with the development of computer science, precise computational prediction based on first-principles theory was becoming possible. Many researchers started to explore prediction models of ceramic materials with low thermal conductivity. Lan etc. [2] studied on the prediction model of dopants effect on pyrochlore ceramic materials and gave a possible prediction range. Malakkal [3], Liu [4], and Sajjad [5] also used first-principles computational methods to predict the thermal conductivity of BeO, $BaZr_2O_3$, and $Cs_2PTi_6$ respectively. In order to enhance the computational efficiency, such as Atomate [6] and FireWorks [7], have been developed in order to automate the tasks assignment and management in computational process. These workflow frameworks effectively realize data exchange between high performance computer cluster and cloud storage and can manage thousands of subtasks. Based on these workflow frameworks, reusable subtask components and computational workflows have been developed for the computation of many material properties including but not limited to elasticity [6,8], dielectric constant [6], and ferroelectricity [9].

Traditional method to calculate lattice thermal conductivities of ceramic materials needs us to obtain 2$^{nd}$ and 3$^{rd}$ force constant of a given unit cell firstly. From 2$^{nd}$ force constant, phonon dispersion and DOS can be further evaluated. And then from 3$^{rd}$ force constant, thermal conductivity can be acquired through solving Boltzmann Transport Equation (BTE). For the gaining of 2$^{nd}$ and 3$^{rd}$ force constants, usually, finite displacement method is used through first-principles calculations (VASP [10]). While there are already open-source libraries like ShengBTE [11], Phonopy [12], and Phono3py [13] assisting forming displacement crystal structure file (POSCAR) and solving Boltzmann equation, the whole process is manually onerous and thus error-prone. And most importantly, these manually repetitive processes cannot fulfill the requirement of high-throughput calculation and screening. Therefore, building an automated workflow for lattice thermal conductivity calculation is quite necessary. Based Debye-Callaway model, a workflow toolkit AICON [14] achieved high-degree of automation and efficiency which excludes the need of calculating the third-order interatomic force constants (IFCs). Although greatly reducing the complexity of workflow components, the accuracy of Debye-Callaway method is not enough to fulfill the precise high-throughput screening. Therefore, developing a workflow based on more accurate model is highly necessary. On top of Atomate and FireWorks, we developed Ph3pyWF, trying to realize an automated workflow for lattice thermal conductivity calculation, in which more flexible user interface and more efficient task management system come true.

# Methodology

## Ph3pyWF workflow

Figure 1 illustrates the schematic diagram of workflow design, including subprocesses and their dependencies involved in a Ph3pyWF workflow. The Ph3pyWF package is developed based on Atomate [6] and FireWorks [7] computational workflow frameworks which allow the defining of workflows with highly reusable subprocesses. Currently, Ph3pyWF supports using VASP [10] code to perform DFT calculations.

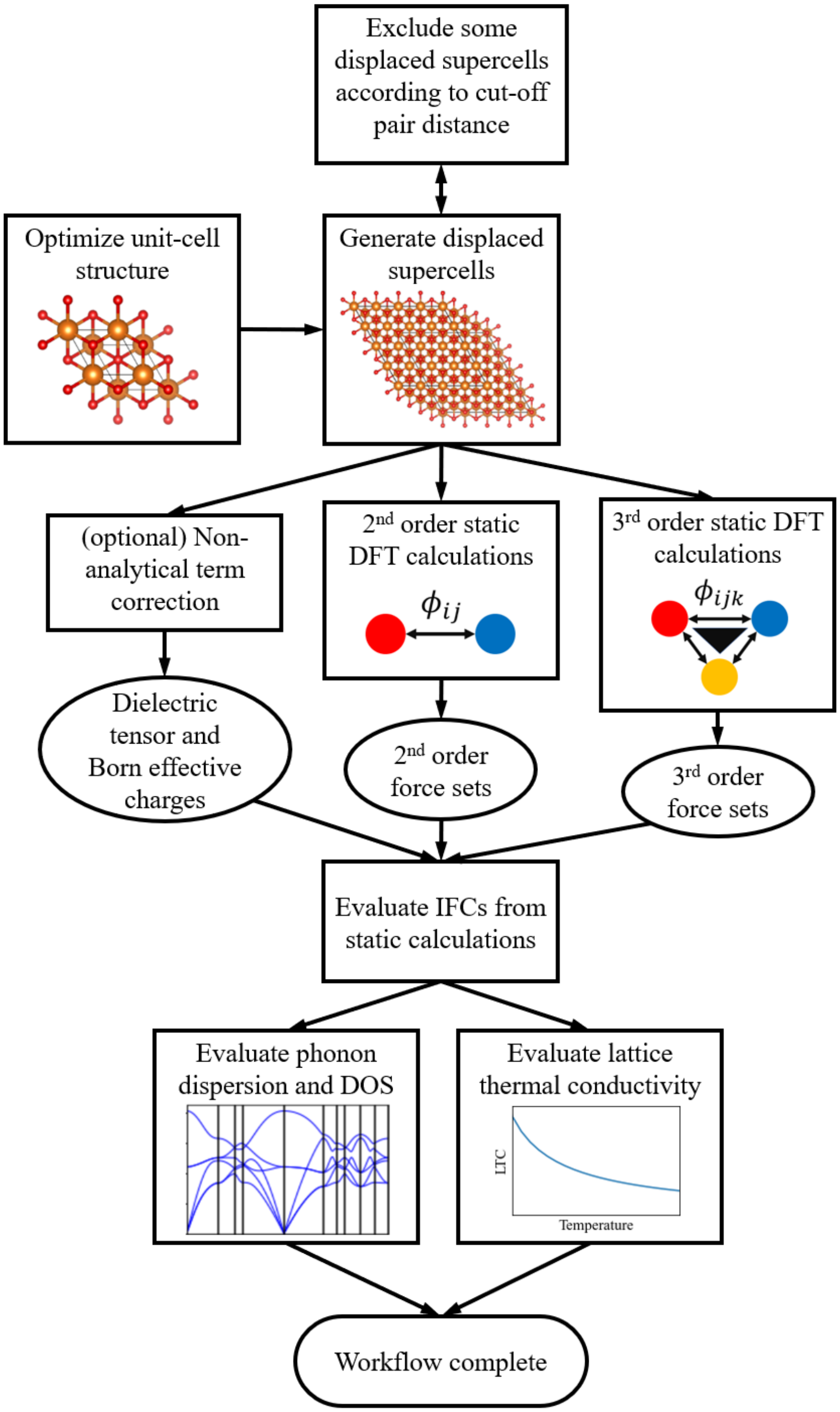

Figure 1: The workflow structure of Ph3pyWF.

The first step of the workflow is to obtain the optimized unit-cell structure of the material system of interest. Preset subprocess for structural optimization available in Atomate library is adopted for this task with slight modification to the default parameters. This subprocess prepares input files for DFT calculation according to the user-specified structure and parameters, runs DFT calculation, and parses output files to extract useful results.

Following the optimized unit-cell structure, displaced supercell structures can then be generated using the finite displacement method. In this step, the optimized unit-cell structure is fetched from the database by querying for the document containing the result of the optimization subprocess. Phono3py [13] is used to generate a list of displaced supercell structures. Users can specify different supercell sizes for the $2^{nd}$ and $3^{rd}$ order IFCs calculation. By default, the $2^{nd}$ order IFCs are calculated from a subset of displaced supercell configurations for the $3^{rd}$ order IFCs. During this subprocess, Phono3py writes a file containing the information of displaced supercell structures, which is later required by the post-analysis subprocess. This file is parsed as a dictionary object to be conveniently stored in the database. Additionally, from a unit-cell with large number of atoms and (or) low symmetry, the number of supercell configurations generated can be exceedingly large which results in unreasonable computation time. Users can specify a cut-off pair distance to exclude some of the supercell configurations. For each displaced supercell structure in the list, a corresponding static DFT job is appended to a list of subprocesses.

Upon the completion of the displaced-supercells-generation subprocess, static DFT subprocesses are dynamically inserted into the workflow as the parents of the post-analysis subprocess. These subprocesses are based on slightly modified preset static calculation subprocess in Atomate library. In the case where different supercell sizes for the $2^{nd}$ and $3^{rd}$ order IFCs calculation are specified, additional subprocesses for the $2^{nd}$ order IFCs are also inserted into the workflow by the aforementioned dynamic insertion. Subprocesses for the $2^{nd}$ IFCs have names formatted differently than those for the $3^{rd}$ order IFCs to provide not only better readability for users, but also more robust data retrieval for post-analysis.

After all the static calculation subprocesses are completed, thermal conductivity can be calculated. The post-analysis subprocess fetches the following data from the database: optimized unit-cell structure, list and information of displaced supercells generated, interatomic forces evaluated by the static calculations, and other required information depending on the specifications. With these sets of data fetched, the subprocess will call the Phono3py to calculate IFCs, and ultimately use the IFCs to solve the Boltzmann Transport equation to get the lattice thermal conductivity at a given range of temperatures. Phonopy library is also called in this subprocess to calculate phonon DOS and dispersion band structure. Post-analysis using Phono3py generates several large output files. The key results (temperature, lattice thermal conductivities, phonon DOS, and dispersion band structure) are stored in the database by default. Optionally, large output files which contain information on phonon group velocities, and phonon frequency can be compressed and stored in the database in binary format.

## Computational parameters

The modular design of this workflow toolkit allows expert users to customize their input parameters at different levels of granularity. **Error! Not a valid bookmark self-reference.** lists the top-most-level parameters used when instantiating a new workflow object. The only required input parameter without default value is the input unit-cell structure, which can be obtained from online database like Materials Project, or by reading files in multiple formats including but not limited to POSCAR and Crystallographic Information File (CIF). If `skip_relax` is set to True, the workflow will skip the structural relaxation and use the input unit cell structure directly to generate displaced supercells. The last parameter `c` is a dictionary object containing more detailed calculation settings. By default, `c` does not contain any data, and any setting specified by the user will override the default calculation settings.

Table 1: Input parameters of Ph3pyWF at workflow level

| Parameter | Type | Default value |
|---|---|---|
| `structure` | pymatgen.Structure | <N/A> |
| `skip_relax` | Boolean | `False` |
| `name` | String | `“phono3py wf”` |
| `c` | Dictionary | `None` |

Table 2 lists the supported parameters in the dictionary object `c`, which would be passed to subprocesses. The parameter `supercell_size_fc3` is a 3-element tuple that specifies the supercell dimension for 3rd order IFCs. Supercell is generated by elongating along lattice axes of unit cell. The default value is (2,2,2), in which case a $2 \times 2 \times 2$ supercell is created. The parameter `supercell_size_fc2` specifies the supercell dimension for 2nd order IFCs and is optional. By default, 2nd order IFCs calculation uses the same supercell dimension as the 3rd order IFCs supercell dimension. Users may specify larger and different supercell dimensions for 2nd order IFCs since two-atom interactions have longer range in real space than three-atom interactions. Once `supercell_size_fc2` is specified, additional displaced supercell structures are generated, and the corresponding static Fireworks will be added to the workflow. It is recommended to set the supercell dimension according to the unit cell structure. A rule-of-thumb is to ensure that the supercell has lengths more than 9 Angstrom along each lattice orientation. The parameter `cutoff_pair_distance` specifies the cutoff pair distance in Angstrom within supercells to reduce the number of static Fireworks. With finite displacement method, two of the three atoms in an interaction triplet are displaced. If the distance between the two displaced atoms in a displaced supercell is larger than the specified cutoff pair distance, such supercell configuration is excluded from the IFCs calculations, and no corresponding static Firework will be added. The parameters `t_min`, `t_max`, and `t_step` specify the temperatures in Kelvin at which lattice thermal conductivities will be evaluated. The default values of `t_min`, `t_max`, and `t_step` generate list temperature values: 200K, 250K, …, 1350K, 1400K. Note that `t_min` is included in the list of temperature values while `t_max` is excluded. The parameter `mesh` is a 3-element list that specifies the q-point mesh sampling grid used for thermal conductivity calculation. The default q-point mesh is 11 points along each axis. For supercell with large number of atoms, $11 \times 11 \times 11$ q-point mesh may result in long computing time exceeding the allocated wall time of a single job, thus it is recommended to specify a less dense q-point mesh initially and run a q-point convergence test.

Table 2: Supported parameters in dictionary object c to modify the calculation settings of subprocesses.

| Parameter | Type | Default value |
|---|---|---|
| tag | String | <Automatically generated> |
| supercell_size_fc3 | Tuple | (2,2,2) |
| supercell_size_fc2 | Tuple | <Same as supercell_size_fc3> |
| cutoff_pair_distance | Float | None |
| atom_disp | Float | 0.03 |
| vasp_input_set_relax | VaspInputSet | Ph3pyRelaxSet |
| vasp_input_set_static | VaspInputSet | Ph3pyStaticSet |
| db_file | String | ">>db_file<<" |
| metadata | Dictionary | {} |
| USER_*_SETTINGS | Dictionary | {} |
| t_min | Float | 200 |
| t_max | Float | 1401 |
| t_step | Float | 50 |
| primitive_matrix | ndarray | None |
| mesh | List | [11,11,11] |
| is_nac | Boolean | True |
| is_symmetry | Boolean | True |
| symprec | Float | 1e-5 |

# Applicability Analysis

To test the applicability of our workflow, some typical binary and complex oxide ceramic structures were tested. Monoclinic $ZrO_2$, cubic MgO, corundum $Al_2O_3$, wurtzite ZnO, wurtzite BeO, rutile $SnO_2$, and cubic $CeO_2$ are selected as binary samples while $La_2Zr_2O_7$ and $Gd_2Zr_2O_7$ are chosen as complex oxides samples. Each system uses a different supercell dimension and a cutoff pair distance to balance accuracy and computational cost.

The lattice thermal conductivities of selected binary oxide systems calculated using Ph3pyWF and their respective experimental results at selected low (300K) and high temperature (900K) are shown in Figure 2. The calculated lattice thermal conductivity values show good agreement with the experimental values, which proves the usability of our workflow in common binary ceramic systems.

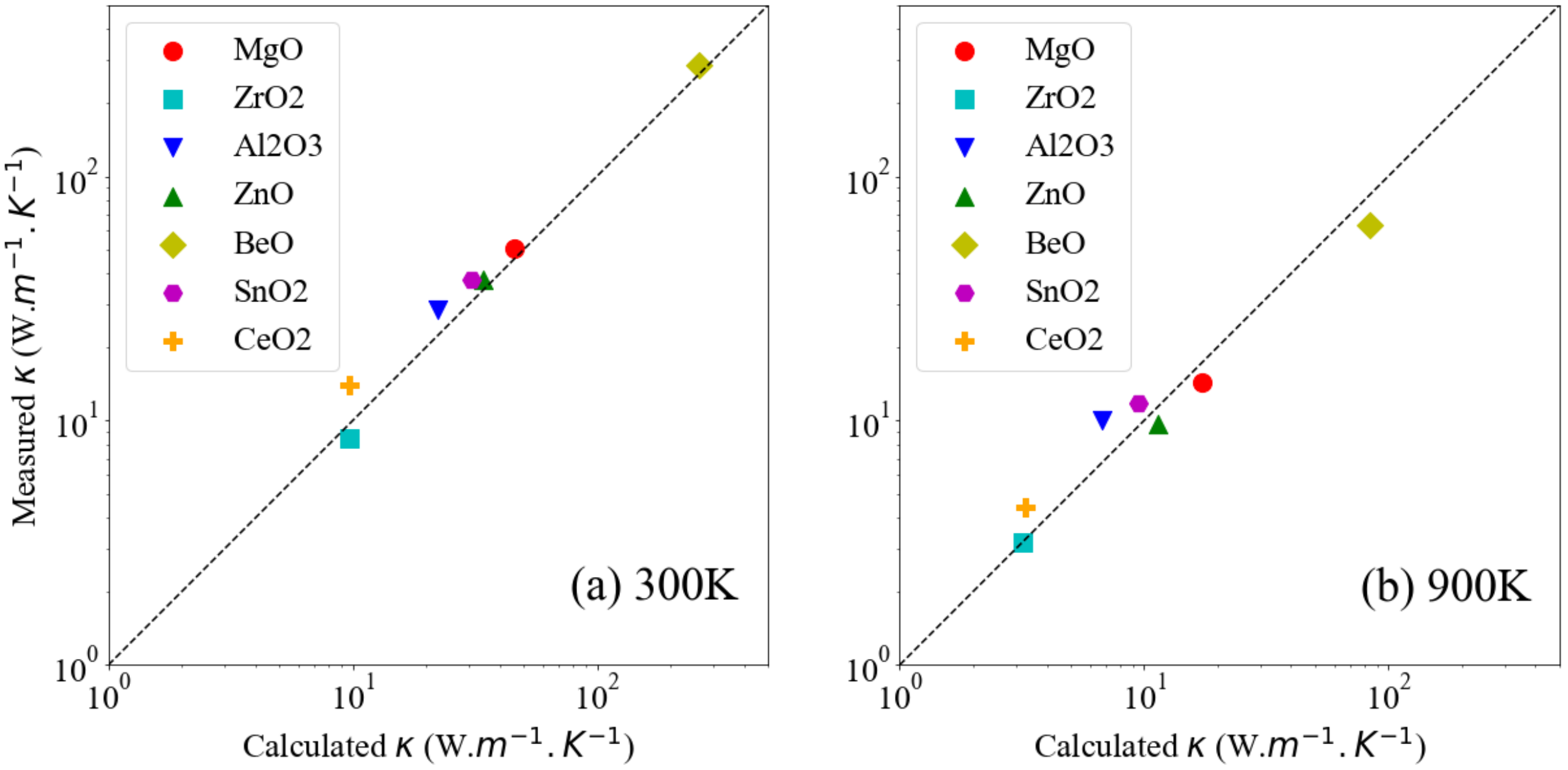

Figure 2: Experimentally measured thermal conductivities of binary oxide systems (MgO [15], ZrO2 [16] , Al2O3 [17], ZnO [18], BeO [19], SnO2 [20], CeO2 [21]) compared with calculated values by Ph3pyWF at (a) low temperature (300K) and (b) high temperature (900K).

Once the validity of the workflow toolkit has been proved within the a few binary oxide systems, we then need to further test Ph3pyWF on more complicated oxide systems such as pyrochlores. Composition of both $La_2Zr_2O_7$ and $Gd_2Zr_2O_7$ are regarded as promising candidates for next generation TBC topcoat materials, which are good as our test samples. The IFCs of $La_2Zr_2O_7$ and $Gd_2Zr_2O_7$ were both calculated using $2 \times 2 \times 2$ supercells (176 atoms) constructed from the primitive cell (22 atoms). A cutoff pair distance of 5 Å was set in both workflows, having reduced the number of static jobs from more than 3800 to less than 900. Local density approximation (LDA) instead of generalized gradient approximation (GGA) was used as exchange-correlation functional to enhance the calculation accuracy in rare-earth system. The lattice thermal conductivities calculated using Ph3pyWF and their experimentally measured results are shown in Figure 3. The calculated results verified the characteristic low lattice thermal conductivities in both systems. For $La_2Zr_2O_7$, the calculated values show good agreement with the experimental ones at relatively low temperature. Underestimation at elevated temperature is observed in both cases, and the flattening pattern at high temperature is not reflected in the calculated lattice thermal conductivities. This can be explained by the significant thermal conductivity contribution from radiational thermal transportation, which is not involved in the calculations of this workflow.

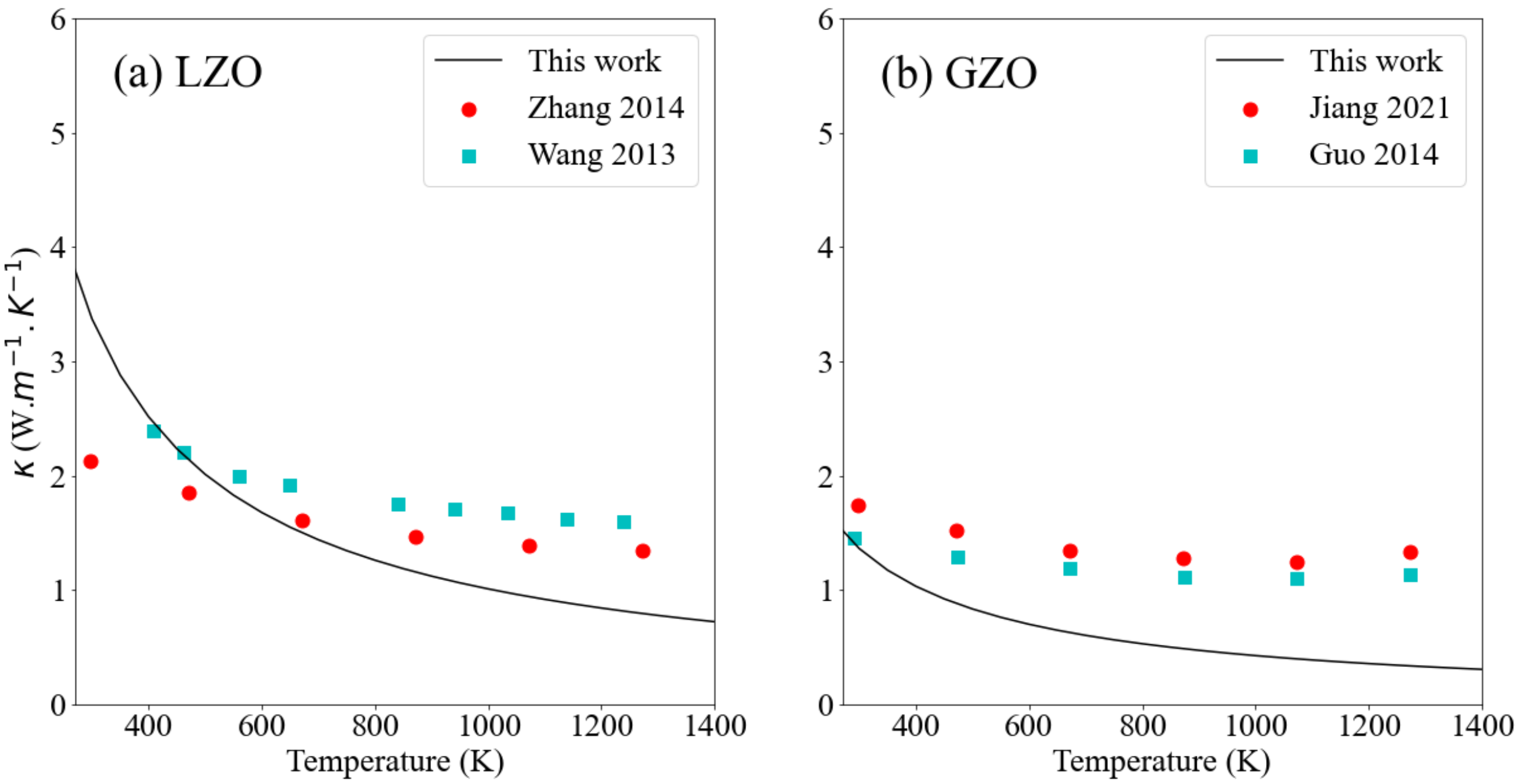

Figure 3: Lattice thermal conductivity ($\kappa$) of (a) pyrochlore $La_2Zr_2O_7$ and (b) pyrochlore $Gd_2Zr_2O_7$. The solid curve represents the calculated values of $\kappa$ using Ph3pyWF. The filled symbols represent the experimental measured values of $\kappa$. [22–25]

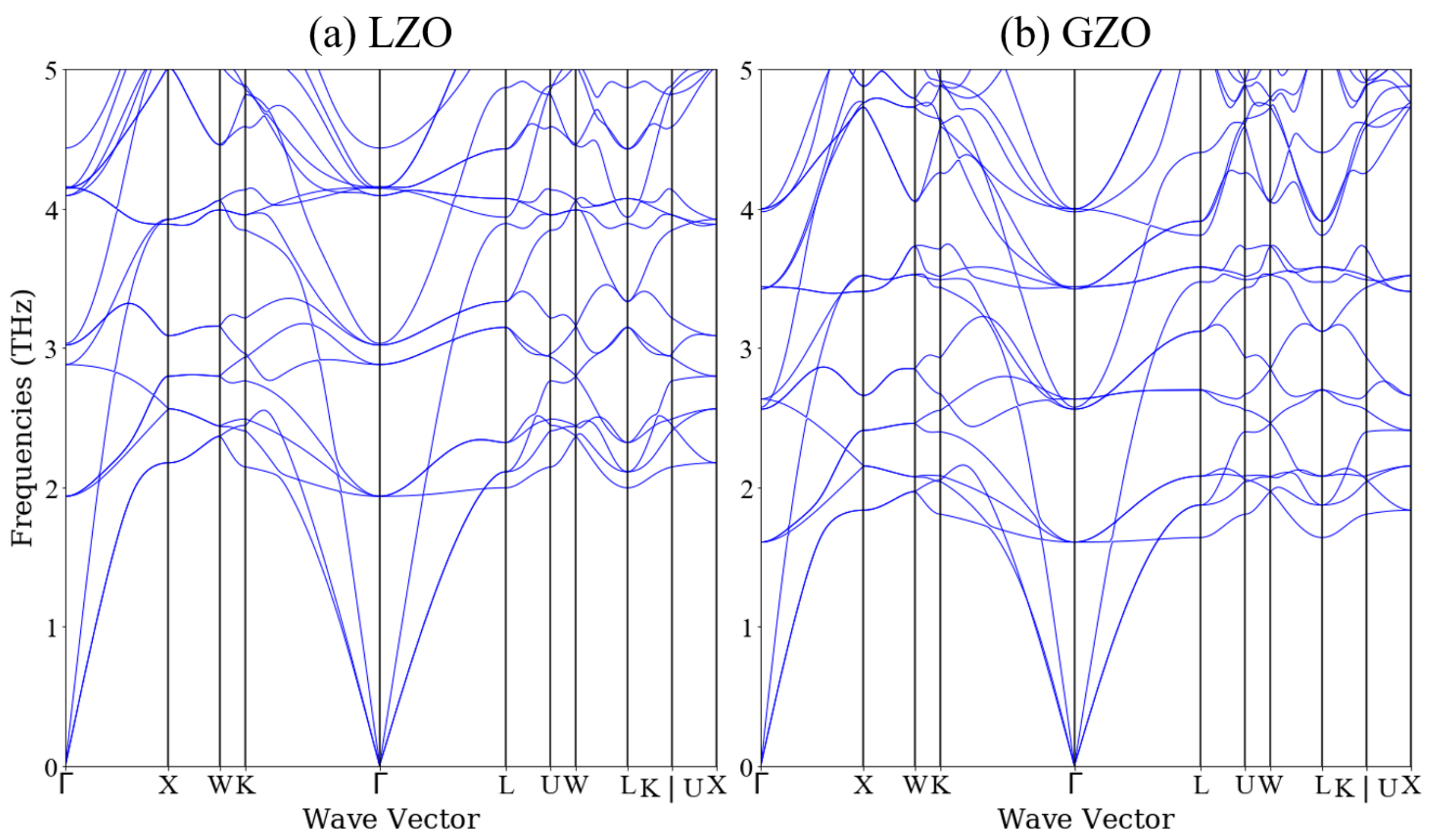

Figure 4: Phonon dispersion band structures for (a) $La_2Zr_2O_7$ and (b) $Gd_2Zr_2O_7$ calculated using Ph3pyWF.

Lan et al. [26] suggested that for pyrochlore systems with rare-earth elements occupying A sites (16d), the lower the frequency of optical phonon branches the lower the thermal conductivity due to the scattering of the transverse acoustic branches caused by the low-lying optical branches. This theory provides reasonable explanation for the lower calculated lattice thermal conductivity of $Gd_2Zr_2O_7$. The calculated results indicate the presence of low-lying optical phonon branches at lower frequency in $Gd_2Zr_2O_7$ (~1.5 THz) than in $La_2Zr_2O_7$ (~2 THz), which can be observed in the phonon dispersion band structures shown in Figure 4.

## Conclusion

In summary, we have presented an automated workflow software package Ph3pyWF for calculating lattice thermal conductivities of TBC materials using the finite displacement method and single-mode relaxation-time (SMRT) approximation. This workflow software employs a state-of-the-art scientific workflow framework to combine multiple computational material science software packages and is expected to accelerate the exploration of new TBC topcoat materials. Ph3pyWF has shown a high degree of automation and is capable of providing a near-turn-key solution for lattice thermal conductivity calculations. Compared with manual procedures, this workflow is highly efficient and avoids error-prone process, therefore suitable for high-throughput screening of materials. The capability of Ph3pyWF has been demonstrated on multiple material systems with within varied temperature range. The results validated the capability and accuracy of this toolkit in predicting the thermal properties of oxides.

## Acknowledgements

The author acknowledges the usage of the Cedar HPC cluster provided by Compute Canada for providing computing resources. This work was supported by National Science and Engineering Research Council of Canada (NSERC) and McGill Engineering Undergraduate Student Masters Award (MEUSMA).

# References

1. Padture, N. P., Gell, M. & Jordan, E. H. Thermal barrier coatings for gas-turbine engine applications. *Science (80-. ).* **296**, 280–284 (2002).
2. Lan, G., Ou, P., Chen, C. & Song, J. A complete computational route to predict reduction of thermal conductivities of complex oxide ceramics by doping: A case study of La2Zr2O7. *J. Alloys Compd.* **826**, 154224 (2020).
3. Malakkal, L., Szpunar, B., Siripurapu, R. K., Zuniga, J. C. & Szpunar, J. A. Thermal conductivity of wurtzite and zinc blende cubic phases of BeO from ab initio calculations. *Solid State Sci.* **65**, 79–87 (2017).
4. Liu, Y. *et al.* Theoretical and experimental investigations on high temperature mechanical and thermal properties of BaZrO3. *Ceram. Int.* **44**, 16475–16482 (2018).
5. Sajjad, M., Mahmood, Q., Singh, N. & Andreas Larsson, J. Ultralow lattice thermal conductivity in double perovskite Cs2PTi6: A promising thermoelectric material. *ACS Appl. Energy Mater.* **3**, 11293–11299 (2020).
6. Mathew, K. *et al.* Atomate: A high-level interface to generate, execute, and analyze computational materials science workflows. *Comput. Mater. Sci.* **139**, 140–152 (2017).
7. Jain, A. *et al.* Fireworks: A dynamic workflow system designed for highthroughput applications. *Concurr. Comput. Pract. Exp.* **27**, 5037–5059 (2015).
8. Wang, Y. *et al.* DFTTK: Density Functional Theory ToolKit for high-throughput lattice dynamics calculations. *Calphad* **75**, 102355 (2021).
9. Smidt, T. E., Mack, S. A., Reyes-Lillo, S. E., Jain, A. & Neaton, J. B. An automatically curated first-principles database of ferroelectrics. *Sci. Data* **7**, 1–22 (2020).
10. Hafner, J. Ab-initio simulations of materials using VASP: Density-functional theory and beyond. *Journal of Computational Chemistry* vol. 29 (2008).
11. Li, W., Carrete, J., Katcho, N. A. & Mingo, N. ShengBTE: A solver of the Boltzmann transport equation for phonons. *Comput. Phys. Commun.* **185**, 1747–1758 (2014).
12. Togo, A. & Tanaka, I. First principles phonon calculations in materials science. *Scr. Mater.* **108**, 1–5 (2015).
13. Togo, A., Chaput, L. & Tanaka, I. Distributions of phonon lifetimes in Brillouin zones. *Phys. Rev. B - Condens. Matter Mater. Phys.* **91**, (2015).

14. Fan, T. & Oganov, A. R. AICON: A program for calculating thermal conductivity quickly and accurately. *Comput. Phys. Commun.* **251**, (2020).
15. Harris, D. C. *et al.* Properties of an infrared-transparent MgO: Y2O3 nanocomposite. *J. Am. Ceram. Soc.* **96**, 3828–3835 (2013).
16. Sun, L., Guo, H., Peng, H., Gong, S. & Xu, H. Phase stability and thermal conductivity of ytterbia and yttria co-doped zirconia. *Prog. Nat. Sci. Mater. Int.* **23**, 440–445 (2013).
17. Parchovianský, M., Galusek, D., Švančárek, P., Sedláček, J. & Šajgalík, P. Thermal behavior, electrical conductivity and microstructure of hot pressed Al2O3/SiC nanocomposites. *Ceram. Int.* **40**, 14421–14429 (2014).
18. Olorunyolemi, T. *et al.* Thermal conductivity of zinc oxide: From green to sintered state. *J. Am. Ceram. Soc.* **85**, 1249–1253 (2002).
19. Takahashi, Y. & Murabayashi, M. Measurement of Thermal Properties of Nuclear Materials By Laser Flash Method. *J. Nucl. Sci. Technol.* **12**, 133–144 (1975).
20. Bueno, P. R., Varela, J. A., Barrado, C. M., Longo, E. & Leite, E. R. A comparative study of thermal conductivity in ZnO- And SnO 2-based varistor systems. *J. Am. Ceram. Soc.* **88**, 2629–2631 (2005).
21. Suzuki, K. *et al.* Thermal and mechanical properties of CeO2. *J. Am. Ceram. Soc.* **102**, 1994–2008 (2019).
22. Zhang, Y. *et al.* Low thermal conductivity in La2Zr2O7 pyrochlore with A-site partially substituted with equimolar Yb2O 3 and Er2O3. *Ceram. Int.* **40**, 9151–9157 (2014).
23. Wang, Y., Yang, F. & Xiao, P. Rattlers or oxygen vacancies: Determinant of high temperature plateau thermal conductivity in doped pyrochlores. *Appl. Phys. Lett.* **102**, (2013).
24. Jiang, T., Xie, M., Guan, L., Wang, X. & Song, X. Effect of Nb5+ and Cu2+ codoping on thermal properties of Gd2Zr2O7 ceramic. *J. Rare Earths* **39**, 180–185 (2021).
25. Guo, L., Guo, H., Peng, H. & Gong, S. Thermophysical properties of Yb2O3 doped Gd2Zr2O7 and thermal cycling durability of (Gd0.9Yb0.1)2Zr2O7/YSZ thermal barrier coatings. *J. Eur. Ceram. Soc.* **34**, 1255–1263 (2014).
26. Lan, G., Ouyang, B. & Song, J. The role of low-lying optical phonons in lattice thermal conductance of rare-earth pyrochlores: A first-principle study. *Acta Mater.* **91**, 304–317 (2015).